\documentclass[pre,aps,10pt,twocolumn]{revtex4-1}
\usepackage{geometry}
\usepackage{graphicx}
\usepackage{amsfonts}
\usepackage{amsmath}
\usepackage{bm}
\usepackage{url}
\usepackage{color}
\usepackage{xfrac}
\usepackage{array,url,kantlipsum}

\newcommand{\ev}{{\bm e}}
\newcommand{\xv}{{\bm x}}
\newcommand{\rv}{{\bm r}}
\newcommand{\vv}{{\bm v}}

\newcommand{\qv}{{\bm q}}
\newcommand{\kv}{{\bm k}}

\newcommand{\Pv}{{\bm P}}
\newcommand{\Av}{{\bm A}}

\newcommand{\tv}{{\bm t}}

\newcommand{\uvkx}{{\hat{\bm k}_x}}
\newcommand{\uvky}{{\hat{\bm k}_y}}
\newcommand{\kpara}{k_ {\parallel}}
\newcommand{\kperp}{k_ {\perp}}

\definecolor{purple}{rgb}{0.5, 0.0, 0.8}
\definecolor{orange}{rgb}{0.9, 0.6, 0.0}

\begin{document}
\title{Hyperuniformity of Quasicrystals} 

\author{Erdal C.~O\u{g}uz}
\affiliation{Department of Chemistry, Princeton University, Princeton, NJ 08540}
\altaffiliation[Present address:\ ]{School of Mechanical Engineering and The Sackler Center for Computational Molecular and Materials Science, 
Tel Aviv University, Tel Aviv 6997801, Israel}
\author{Joshua E.~S.~Socolar}
\affiliation{Department of Physics, Duke University, Durham, NC 27708}
\author{Paul J.~Steinhardt}
\affiliation{Princeton Center for Theoretical Science and Department of Physics, Princeton University, Princeton, NJ 08544}
\author{Salvatore Torquato}
\affiliation{Department of Chemistry, Department of
Physics, Princeton Institute for the Science and Technology of Materials, and Program in Applied and Computational
Mathematics, Princeton University, Princeton, 08540}

\date{\today}

\begin{abstract}
Hyperuniform systems, which include crystals, quasicrystals and special disordered systems, have attracted
considerable recent attention, but rigorous analyses of the hyperuniformity of quasicrystals have been lacking because the support of the spectral intensity is dense and discontinuous. We employ the integrated spectral intensity, $Z(k)$, to quantitatively characterize the hyperuniformity of quasicrystalline point sets generated by projection methods.  The scaling of $Z(k)$ as $k$ tends to zero is computed for one-dimensional quasicrystals and shown to be consistent with independent calculations of the variance, $\sigma^2(R)$, in the number of points contained in an interval of length $2R$.  We find that one-dimensional quasicrystals produced by projection from a two-dimensional lattice onto a line of slope $1/\tau$ fall into distinct classes determined by the width of the projection window.  For a countable dense set of widths, $Z(k) \sim k^4$; for all others, $Z(k)\sim k^2$. This distinction suggests that measures of hyperuniformity define new classes of quasicrystals in higher dimensions as well.
\end{abstract}

\maketitle 

\section{Introduction}
Hyperuniform many-particle systems have density fluctuations that are anomalously suppressed at long wavelengths compared to the fluctuations in typical disordered point configurations, such as atomic positions in ideal gases, liquids and glasses.  For disordered systems, a hyperuniform many-particle system in $d$-dimensional Euclidean space $\mathbb{R}^d$ is one in which the structure factor $S({\kv})$ tends to zero as the wavenumber $k\equiv |\kv|$ tends to zero \cite{To03a}; i.e.,
\begin{equation}
\lim_{|{\kv}| \rightarrow 0} S({\kv}) = 0.
\label{eqn:hyper}
\end{equation}
Equivalently, it is one in which the local number variance of particles within a spherical observation window of radius $R$, denoted by $\sigma^2(R)$,  grows as $R^{\nu}$ in the large-$R$ limit with $\nu<d$ in $d$ dimensions. Typical disordered systems, such as liquids and structural glasses, have the standard volume scaling  $\sigma^2(R) \sim R^d$.  By contrast,  for perfect crystals  the variance grows only like the surface area $\sigma^2(R)\sim R^{d-1}$, making them hyperuniform \cite{To03a,Za09}.  There are various classes of disordered particle configurations that are hyperuniform, and their novel structural and physical properties have received considerable recent attention \cite{Ba08,Fl09b,Man13b,Ha13,Ji14,Zi15,Le16}.  Numerical calculations have also demonstrated that certain quasicrystalline point sets have $\sigma^2(R) \sim R^{d-1}$ and hence are hyperuniform~\cite{Za09,Fl09b}.  
It is also known that other one-dimensional quasicrystalline point sets, while still hyperuniform, show a logarithmic growth in $\sigma^2(R)$ \cite{Kuipers1974,Aubry1987,Aubry1988}.
 For quasicrystalline systems, however, Eq.~(\ref{eqn:hyper}) requires reconsideration because $S(\kv)$ is everywhere discontinuous, being comprised of a dense set of Bragg peaks \cite{Le84}.

There is a deep connection between the scaling of the local number variance $\sigma^2(R)$ and the behavior of $S(\kv)$ for small $|\kv|$~\cite{To03a}.  For a general point configuration with a well-defined average number density,  $\sigma^2(R)$ is determined entirely by pair correlations and can be expressed in terms of $S(\kv)$ and the Fourier transform ${\tilde \mu}(k;R)$ of a uniform density sphere of radius $R$:
\begin{equation}
\sigma^2(R)=
\rho v_1(R)\Bigg[\frac{1}{(2\pi)^d} \int_{\mathbb{R}^d} S({\kv}) {\tilde \mu}(k;R) d{\kv}\Bigg] 
\label{eqn:local}
\end{equation}
with 
\begin{equation}
{\tilde \mu}(k;R)= 2^d \pi^{d/2} \Gamma(1+d/2)\frac{[J_{d/2}(kR)]^2}{k^d}\,,
\label{eqn:alpha-k}
\end{equation}
where $\rho$ is the density, 
$v_1(R)= \pi^{d/2} R^d/\Gamma(1+d/2)$ is the volume of a $d$-dimensional spherical window,
the wavenumber $k$ is the magnitude of $\kv$, and
$J_{\nu}(x)$ is the Bessel function of order $\nu$.

In cases where the structure factor goes to zero continuously as 
\begin{equation}
S({\kv}) \sim k^\alpha \quad (\alpha>0)\,,
\label{eqn:power}
\end{equation}
it follows from Eq.~(\ref{eqn:local}) that the number variance has the following large-$R$ asymptotic scaling \cite{To03a,Za09,Za11b}:
\begin{equation} \label{eqn:alphanu}
\sigma^2(R) \sim \left\{\begin{array}{ll}
R^{d-1}, & \alpha > 1 \\
R^{d-1}\ln R, & \alpha = 1 \\
R^{d-\alpha}, &  \alpha < 1 
\end{array}\right. \quad R\rightarrow\infty\,.
\end{equation}
We use the term {\it strongly hyperuniform} to refer to systems exhibiting the minimal variance scaling exponent $\nu = d-1$.

Perfect crystals with a finite basis have $S(\kv)=0$ for all $k$ smaller than the first Bragg peak in reciprocal space, which may be interpreted as  
corresponding to the limit $\alpha \rightarrow \infty$.  Maximally random jammed (MRJ) sphere packings \cite{Do05d}, as well as the ground states of free fermions \cite{To08c}
and of superfulid helium \cite{Fe56,Re67}, have $\alpha=1$; one-component plasmas and randomly perturbed lattices have $\alpha=2$; and
certain classical potential energy functions possessing disordered ground states can be tuned so that $\alpha$ can take any positive value \cite{Uc06b,Za11b}.
Note that Eqs.~(\ref{eqn:hyper}) and~(\ref{eqn:power}) assume that the magnitude of the structure factor as the wavenumber goes to zero  is
independent of the wave vector direction.  This standard definition of hyperuniformity has recently been generalized to account for anisotropic spectral functions~\cite{To16a}.
One advantage of the reciprocal-space hyperunformity definition is that it is a property of the point set itself, whereas the behavior of $\sigma^2(R)$ for large $R$ can depend on the choice of window shape \cite{Kim2016}.

A challenge in interpreting Eq.~(\ref{eqn:power}) arises for cases in which the structure factor is {\it discontinuous with dense support} or strongly {\it singular} for arbitrarily small $k$.  Well-known examples are quasicrystals and incommensurate crystals, for which $S(\kv)$ consists of a dense set of Bragg peaks separated by gaps of arbitrarily small size \cite{Le84}. For example, for one-dimensional (1D) quasicrystals, $S(k)$ consists of $\delta$-functions at $k = 2\pi (p + q \tau)/\ell$ for all integers $p$ and $q$ and an irrational value of $\tau$, with $\ell$ being the average spacing between points.  This means that there are peaks arbitrarily close to $k = 0$, and a new, robust criterion to identify and characterize hyperuniformity in such systems is required.

In this paper, we identify an improved hyperuniformity criterion that matches the earlier definitions for crystals and systems with continuous $S(\kv)$ but also serves to characterize quasicrystals and other structures with discontinuous $S(\kv)$.  
The new metric arises from the simple observation that Eq.~(\ref{eqn:local}) has, after integration by parts, the alternative representation
\begin{equation}
\sigma^2(R)=
-\rho v_1(R)\Bigg[\frac{1}{(2\pi)^d} \int_0^\infty  Z(k) 
\frac{\partial {\tilde \mu}(k;R)}{\partial k} dk \Bigg]  ,
\label{eqn:local-1}
\end{equation}
where
\begin{equation}
Z(k) = \int_0^k  S({\qv}) s_d\, q^{d-1}dq
\label{Z}
\end{equation}
is the {\it integrated} or {\it cumulative} intensity function within a sphere of radius $k$
of the origin in reciprocal space, and 
$s_d =d\, \pi^{d/2}/\Gamma(1+d/2)$ is the surface area of a $d$-dimensional sphere of unit radius. 
For simplicity, we have assumed here an isotropic system, but this restriction is easily relaxed.

The fact that the cumulative intensity function $Z(k)$ is smoother than $S(\kv)$
can be exploited to extract the value of $\alpha$ appearing in Eq.~(\ref{eqn:alphanu}), even when 
$S(\kv)$ consists of dense Bragg peaks.
As we shall see below, for quasicrystals $Z(k)$ is a monotonic function with the property
\begin{equation}
c_- k^{\alpha +1} < Z(k)< c_+ k^{\alpha +1} 
\end{equation}
for some constants $c_-$ and $c_+$, some value of $\alpha$, and sufficiently small $k$.
As shorthand for this condition, we say 
\begin{equation}
Z(k) \sim k^{\alpha +1} \quad {\rm as} \; k \rightarrow 0
\label{eqn:Z2}
\end{equation}  
though, strictly speaking, the limit may not exist because $Z(k)$ is an oscillatory function of $\log(k)$.
As before, hyperuniformity corresponds to $\alpha >0$.  The value of $\alpha$ obviously agrees with the previous definition for cases where $S(\kv)$ is a smooth function, since the former is obtained by differentiating the cumulative intensity $Z(k)$ with respect to $k$.  

In the remainder of this paper, we focus exclusively on one-dimensional quasicrystals produced via the standard projection method, in which a subset of points of a two-dimensional lattice is projected onto a line whose slope is incommensurate with that lattice.  We show here that extracting $\alpha$ from $Z(k)$ leads to values consistent with Eq.~(\ref{eqn:alphanu}) for quasicrystalline point sets.  Because the original lattice, being a crystal, is strongly hyperuniform and the subset is determined by taking all points within a uniform width strip parallel to the projection line, one might intuitively expect the values of $\alpha$ for the resulting quasicrystal to correspond to strong hyperuniformity as well.  We find, however, that there are two classes of quasicrystals with different values of $\alpha$, one of which does not conform to the expectation of strong hyperuniformity.
 
We note that there are also 1D structures with more exotic forms of $Z(k)$ than those treated here, such as tilings produced by projections from higher dimensions or by substitution rules.  The latter will be addressed in a separate publication.

\section{Quasicrystals generated by projection}
\label{sec:projection}
\begin{figure}[b]
\centering
\includegraphics[width=7.0cm]{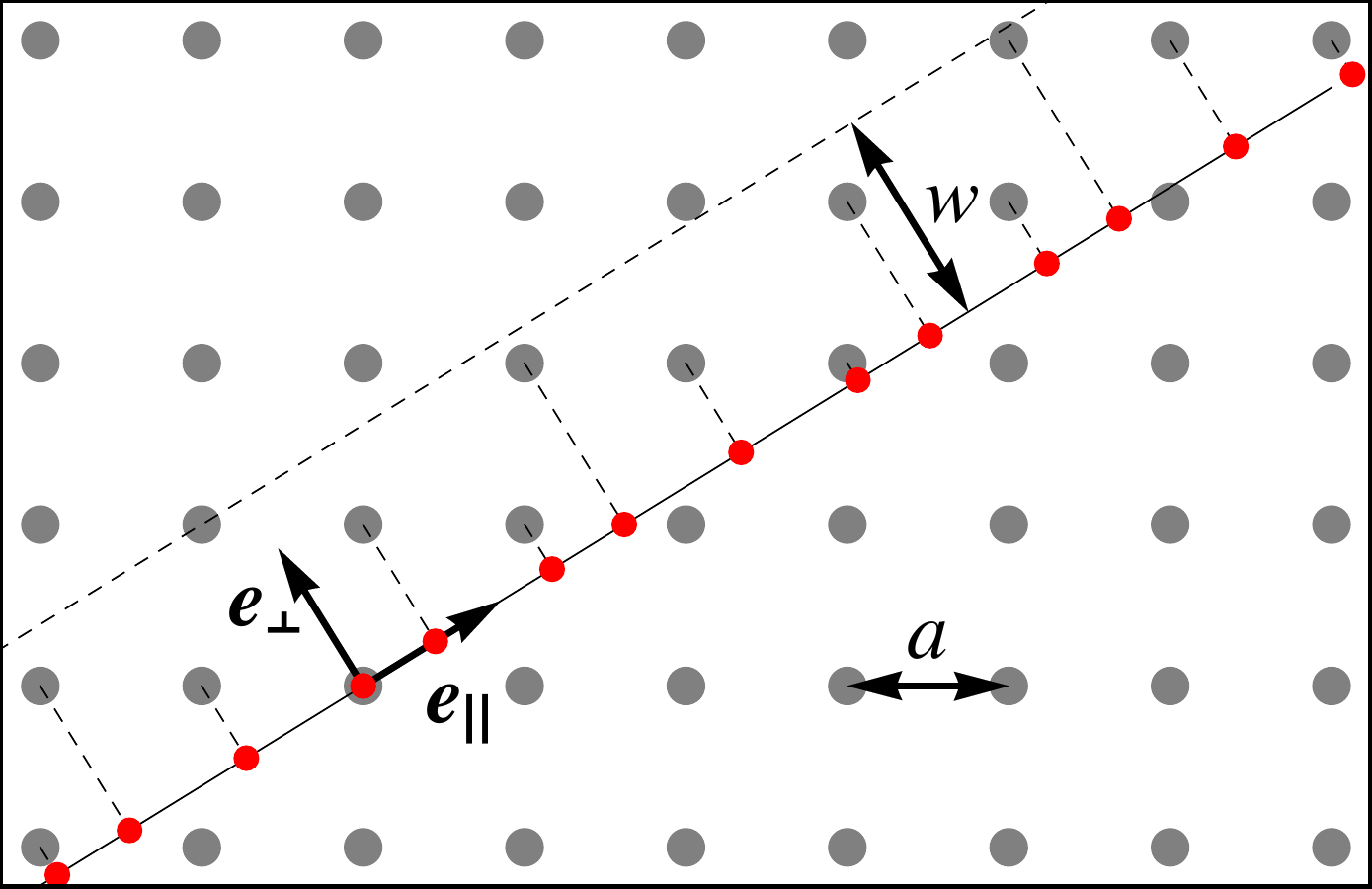}
\caption{Projection of lattice points to create the 1D point set of interest.  The red dots lie in the physical space $X$.}
\label{fig:projection}
\end{figure}

We consider point sets obtained from projection of a subset points of 2D square lattice onto a line of slope $1/\tau$, called the {\em physical space}.  The points selected for projection are those whose orthogonal projections onto the {\it perp-space}, the orthogonal complement of the physical space, lie within a fixed segment of length $w$.  In other words, the lattice points chosen for projection lie within an infinite strip of width $w$ oriented parallel to the physical space, as shown in Fig.~\ref{fig:projection}.  For technical reasons, we specialize to the case $\tau = \left(1+\sqrt{5}\right)/2$, the golden ratio.  We refer to the projected point sets as ``Fibonacci quasicrystals.''  The generalization to $\tau$ of the form $\left(m + \sqrt{m^2 + 4}\right)/2$ for any integer $m$ is straightforward.

\subsection{$Z(k)$ and the scaling exponent $\alpha$}
We begin by computing the structure factor $S(k)$ for the projected tiling.  It is convenient to define a dimensionless measure, $\omega$, of the width of the projection strip by setting the width $w$ equal to $a\,\tau\,\omega / \sqrt{1+\tau^2}$, where $a$ is the lattice constant of the 2D lattice. The calculation, explained in Appendix~\ref{app:Sk}, yields the following result:
\begin{equation}\label{eqn:generalSkpq}
S(k_{pq}) = C' \left(\frac{(p+q\tau)\,\sin\left[\pi\omega \left(p - \frac{p+q\tau}{1+\tau^2}\right)\right]}{p^2-q^2+p\,q}\right)^2\,,
\end{equation}
where $C'$ is a constant independent of $p$ and $q$.

For notational convenience, we define 
\begin{equation}
k_{pq}  \equiv \frac{2\pi(p + q\tau)}{a\sqrt{1+\tau^2}}
\end{equation}
and
\begin{equation} \label{eqn:Ipq}
I_{pq}  \equiv |p^2 - q^2 + p\,q|\,.
\end{equation}
Multiplication of $k_{pq}$ by $1/\tau$ yields $k_{p'q'}$ with $p' = q-p$ and $q'=p$.  Under this operation, $I_{pq}$ is invariant, so we can organize the peaks into sequences with simple scaling properties in the low-$k$ limit.  

\begin{figure}
\centering
\includegraphics[width=0.8\columnwidth]{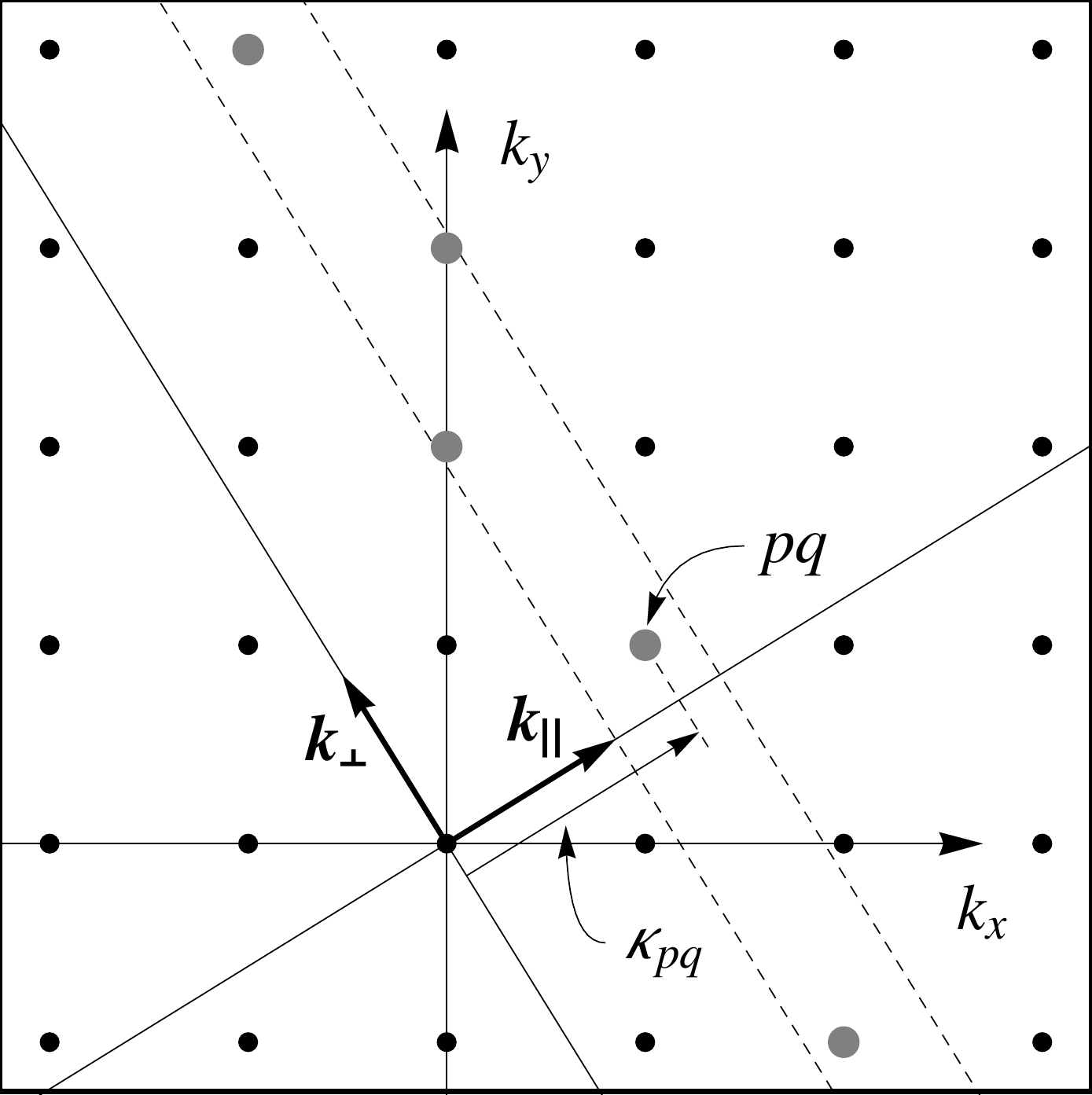}
\caption{Scaling classes in reciprocal space for the Fibonacci projection tilings.  Each large gray dot belongs to a distinct scaling class labeled by $pq$, 
with $\kappa_{pq}$ being the wavenumber of the element of that class lying between $2\pi/a$ and $2\pi\tau/a$ (dashed lines). 
}
\label{fig:kappapq}
\end{figure}

Let $k_n$ denote the scaling sequence $k_{pq}/\tau^n$ where $n = 0, 1, 2, \ldots$, and note that the region $2\pi/a \leq k_{pq} < 2\pi\tau/a$ contains exactly one peak in each scaling sequence.  We let $\kappa_{pq}$ designate these peak positions, as shown in Fig.~\ref{fig:kappapq}.  

To extract the behavior of $S(k)$ for a given scaling sequence, care must be taken with the argument of the sine function in Eq.~(\ref{eqn:generalSkpq}).  We refer to windows corresponding to choices of $\omega$ of the form $i+j/\tau$ with integer $i$ and $j$ as ``ideal windows.''  For an ideal window, the argument of the sine can then be written as
\begin{equation}
\pi\left(ip - jq + \left[\frac{j}{\tau}-\frac{\omega}{1+\tau^2}\right]\epsilon_{pq}\right)\,,
\end{equation}
where $\epsilon_{pq} \equiv p + q\tau$ (which is proportional to $k_{pq}$), and we have used the identity $(j/\tau)p = (j/\tau)\epsilon_{pq} - jq$.    
The integer multiples of $\pi$ have no effect on the magnitude of the sine, so we may rewrite Eq.~(\ref{eqn:generalSkpq}) as
\begin{equation} \label{eqn:Skpq}
S(k_{pq}) = \frac{C'}{I_{pq}^2}\left(\epsilon_{pq}\,\sin\left[\pi\epsilon_{pq}\left(\frac{j}{\tau}-\frac{\omega}{1+\tau^2}\right)\right]\right)^2\,.
\end{equation}

\begin{figure}
\centering
\includegraphics[width=0.8\columnwidth]{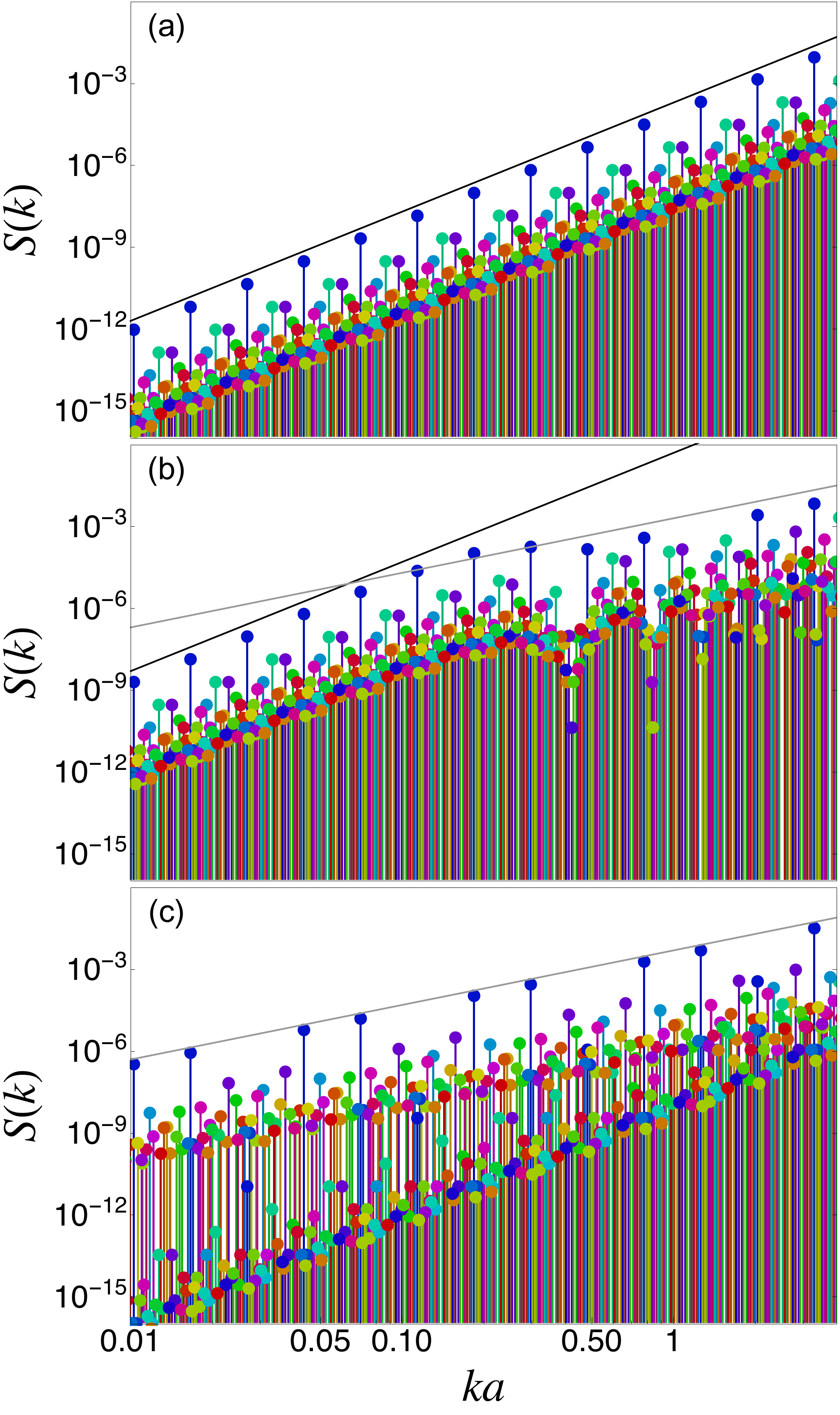}
\caption{Scaling of $S(k)$ at small $k$ for Fibonacci projection tilings constructed from different window widths.  The scaling sequence associated with a 15 smallest values of the invariant $I_{pq}$ are shown, each in a different color.  Black and gray lines have slope $4$ and $2$, respectively. (a) The canonical case $\omega = 1+1/\tau$.  (b) $\omega = 9 - 12/\tau$.  (c)  $\omega = \tau/2$, for which the window is not ideal.}
\label{fig:scaling}
\end{figure}

For any given $j$ and $\omega$, the argument of the sine in Eq.~(\ref{eqn:Skpq}) approaches zero as $\epsilon_{pq}$ approaches zero, and the peak intensities scale like $\epsilon_{pq}^4$, or $k_n^4$.  As one might expect, for larger strip widths (larger $\omega$), the quartic scaling sets in at smaller values of $k_n$ because the density of the system is larger and the entire spectrum is compressed.  More surprisingly, the crossover from quadratic to quartic scaling can set in at very different values of $k_n$ for strips of nearly equal width due to the fact that expressing $\omega$ in terms of $i$ and $j$ may require vastly different values of $j$.  Figures~\ref{fig:scaling}(a) and (b) show examples of $S(k)$ for $\omega = 1+1/\tau = 1.61803\ldots$ and $9-12/\tau = 1.58359\ldots$, with intensities determined from Eq.~(\ref{eqn:generalSkpq}).

If $\omega$ is a real number {\em not} of the form $i+j/\tau$, closer approximations of $\omega$ require ever larger values of $j$, making $j$ effectively infinite.  Thus $j \epsilon_{pq}$ is never small, and the sine function continues to oscillate as $k_n$ approaches zero.  The crossover to quartic scaling never occurs, and the scaling is determined only by the factor of $\epsilon_{pq}$ outside the sine function, leading to $S(k)\sim k_n^2$.  An example is shown in Fig.~\ref{fig:scaling}(c), where $\omega = \tau/2$. 

To compute $\alpha$, the hyperuniformity scaling exponent defined by Eq.~(\ref{eqn:Z2}), we need to show that $Z(k)$ is bounded both above and below by functions of the form $c_{\pm}\,k^{1+\alpha}$ for small $k$.
Within a scaling sequence labeled ``$pq$'', the Bragg peak intensities at $k_n = \kappa_{pq} /\tau^n$ scale as $(1/I_{pq}^2) k_n^{\gamma}$ 
for sufficiently large $n$, where $\gamma$ is the exponent characterizing the envelope of $S(k)$.  The largest $k_n$ that is smaller than $k$ corresponds to $n \equiv n_{pq} = \lceil\ln (\kappa_{pq}/k)/\ln \tau\rceil$, where $\lceil x\rceil$ is the smallest integer greater than $x$.  To get $Z(k)$, we must sum the intensities of all peaks with $n\ge n_{pq}$ in each scaling sequence.

\begin{figure}
\centering
\includegraphics[width=0.8\columnwidth]{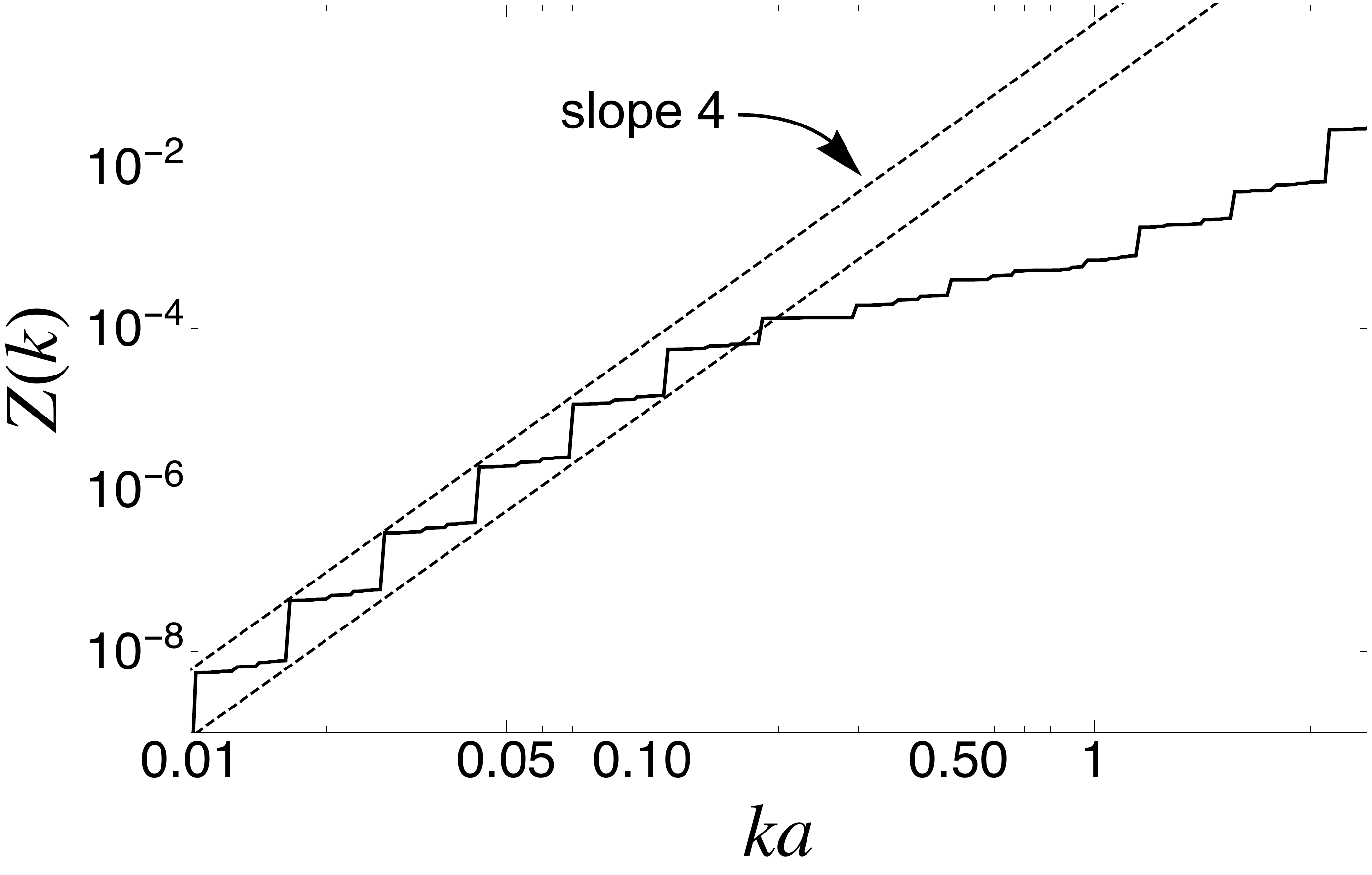}
\caption{Behavior of $Z(k)$ for a Fibonacci projection tiling computed by direct summation of the peak intensities in Fig.~\ref{fig:scaling}(b) .  Dashed lines indicate predicted upper and lower bounds with $c_+ = 0.6$ and $c_-=0.6 \tau^{-4}$.  The curve lies within these bounds for sufficiently small $k$.  The scaling exponent $\gamma = 4$ is in the strongly hyperuniform range.}
\label{fig:Z}
\end{figure}

We first treat the case of ideal windows: $\omega=i+j/\tau$.  Here the argument of the sine in Eq.~(\ref{eqn:Skpq}) approaches zero for large $n$ for any given $j$ and $\omega$.  Thus the sine function differs from its argument only by terms of order $\epsilon_{pq}^2$.  Recall that $\gamma = 4$ for this case.
We have
\begin{align}
Z(k) & =  C'' \sum_{pq} \sum_{n=n_{pq}}^{\infty} \frac{1}{I_{pq}^2}\,\left(\frac{\kappa_{pq}^{\gamma}}{\tau^{\gamma n}}\right) +{\cal O}(\tau^{-2 \gamma n_{pq}}) \nonumber \\
\xrightarrow[{\rm large\ }n_{pq}]{} & \, C''\left(\frac{1}{1-1/\tau^{\gamma}}\right)\sum_{pq}  \frac{1}{I_{pq}^2}\left(\frac{\kappa_{pq}}{\tau^{n_{pq}}}\right)^{\gamma} \nonumber \\ 
\  & <  C'' \left(\frac{k^{\gamma}}{1-1/\tau^{\gamma}}\right)\sum_{pq}  \frac{1}{I_{pq}^2} \,,
\label{eqn:boundZ}
\end{align}
where the sums over $pq$ are taken over the distinct scaling classes and 
\begin{equation}
C'' = C' \pi \left(\frac{j}{\tau}-\frac{\omega}{1+\tau^2}\right)\,.
\end{equation} 
The inequality in the last line of Eq.~(\ref{eqn:boundZ}) is due to the fact that $\kappa_{pq}/\tau^{n_{pq}} < k$, with the possible exception of a single point if $k = \kappa_{pq}$ for some $pq$.  For the gray dots in Fig.~\ref{fig:kappapq}, we have $q \approx -p\tau$ for large $p$ and hence $I_{pq}\sim 2\tau p^2$, so the sum over $pq$ class invariants converges. 
Thus we have shown that $Z(k)$ is bounded above by $c_+ k^{\gamma}$, with $c_+ = C' (1-1/\tau^{\gamma})^{-1} \sum I_{pq}^{-2}$.  Noting that $k/\tau < \kappa_{pq}$, the same reasoning applies but now with the inequality reversed and an additional factor of $\tau^{-\gamma}$ on the right hand side, establishing that $Z(k)$ is bounded below by $c_- k^{\gamma}$, with $c_- = c_+/\tau^{\gamma}$.  Figure~\ref{fig:Z} shows $Z(k)$ and the derived upper and lower bounds for the system of Fig.~\ref{fig:scaling}(b).

\begin{figure}
\centering
\includegraphics[width=0.8\columnwidth]{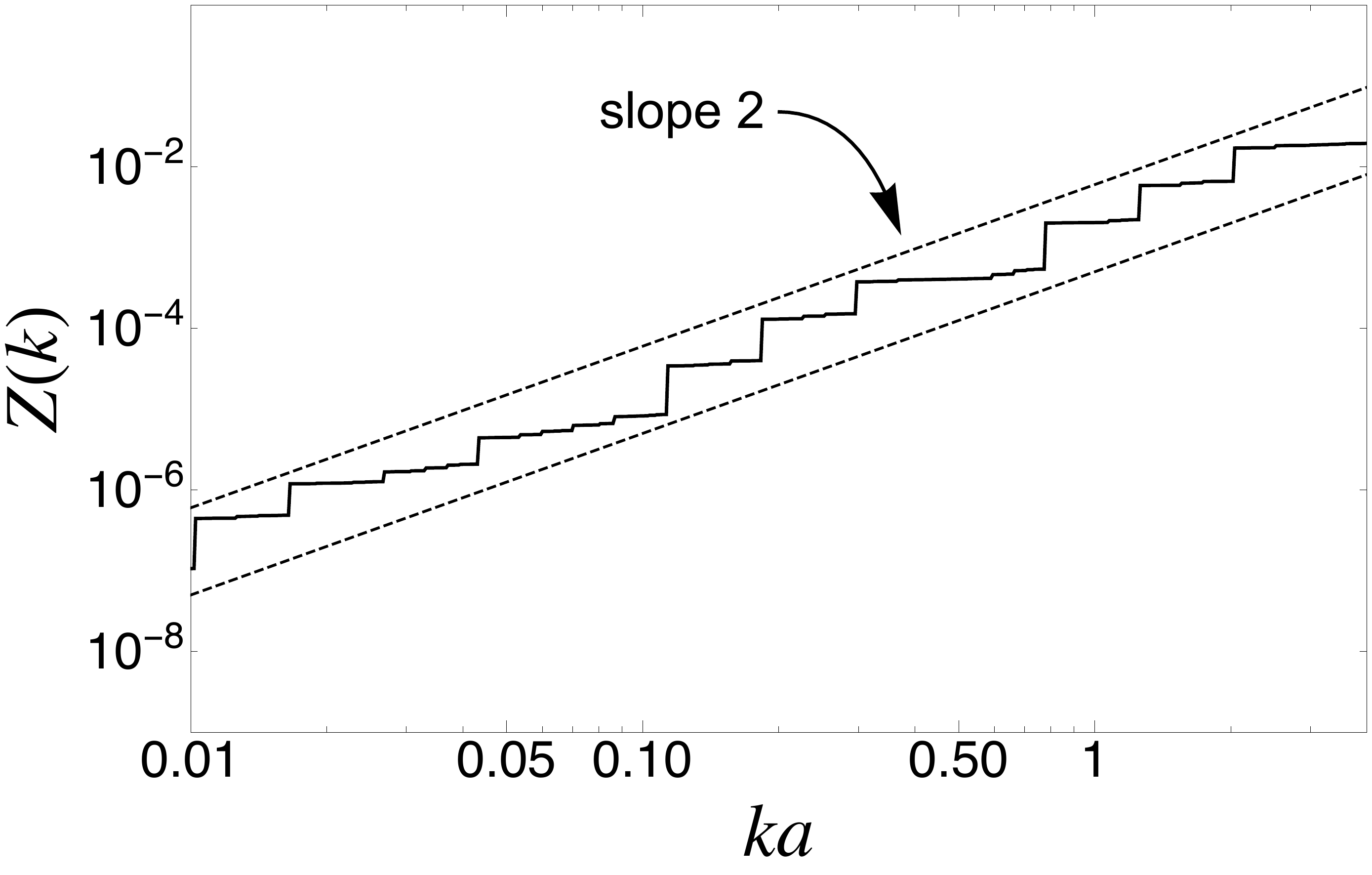}
\caption{Behavior of $Z(k)$ for a Fibonacci projection tiling computed by direct summation of the intensities of the 15 strongest scaling sequences for $\omega = \sqrt{2}$.  Dashed lines indicate derived upper bound and apparent lower bound with exponent $\gamma=2$.  Note that this exponent is smaller than that of Fig.~\ref{fig:Z} and is not correspond to strong hyperuniformity.}
\label{fig:Znonideal}
\end{figure}

For non-ideal $\omega$, the argument of the sine in Eq.~(\ref{eqn:generalSkpq}) approaches $p_n\pi\omega$ for large $n$, which does not converge to zero.  Recall that $\gamma = 2$ for this case.  An upper bound on $Z(k)$ is easily obtained by setting the sine to unity, immediately yielding
\begin{align}
Z(k) & <  C' \sum_{pq} \sum_{n=n_{pq}}^{\infty} \frac{1}{I_{pq}^2}\,\left(\frac{\kappa_{pq}^{\gamma}}{\tau^{\gamma n}}\right)  \nonumber \\
\  & <  C' \left(\frac{k^{\gamma}}{1-1/\tau^{\gamma}}\right)\sum_{pq}  \frac{1}{I_{pq}^2} \,,
\label{eqn:boundZ2}
\end{align}
The lower bound is more difficult to establish because the sine jumps erratically with $n$ and can take on values arbitrarily close to zero for some terms.  
When $\omega$ is a rational multiple of some $i+j/\tau$, the values of the sine in any given scaling sequence converge to a periodic variation with $n$, as is readily visible in Fig.~\ref{fig:scaling}(c).  In such cases, one can always identify subsequences of the scaling sequence for which the sum entering $Z(k)$ scales like $k^{\gamma}$, which is sufficient to establish that the full $Z(k)$ must scale like $k^{\gamma}$ and, in fact, the above derivation of $c_-$ provides a tighter bound.  When $\omega$ is not rationally related to any number of the form $i+j/\tau$, this argument cannot be applied, and we do not have a rigorous proof of the lower bound.  Numerical evidence strongly suggests, however, that there is such a bound.  An example is shown in Fig.~\ref{fig:Znonideal}.

We have thus established that $Z(k)$ scales like $k^{\gamma}$ for sufficiently small $k$.  For the case of generic window width ($\gamma=2$), this gives $\alpha = 1$, while for ideal windows ($\gamma=4$) we have $\alpha = 3$.

\subsection{Calculation of the number variance $\sigma^2(R)$}

For quasiperiodic 1D sequences, the distribution of the numbers of points within segments of a given finite length has been studied extensively as a topic in discrepancy theory \cite{Kesten1966,Kuipers1974,Beck2001}. The results reported here, together with Appendices B and C, are consistent with previously obtained results for closely related sequences.

We show here that the values of $\alpha$ we have obtained are consistent with direct calculations of $\sigma^2(R)$.  For $\omega$ of the form $i+j/\tau$, $\sigma^2(R)$ can be computed analytically for all $R$.  For the generic case, we develop a double sum over hyperlattice reciprocal space vectors that can be numerically evaluated.  The calculations of $\sigma^2(R)$ apply to projections onto a line of arbitrary slope.  Our treatment here is general, so we use the symbol $\beta$, with the Fibonacci case corresponding to $\beta = \tau$.

When $\omega$ is of the form $i+j/\beta$ and the projection strip is positioned such that its lower boundary passes through the origin of the 2D lattice, the width of the projection strip $w$ is such that the upper boundary also passes through some lattice point $\vv$.  The lower boundary is assumed to be closed, while the upper boundary is taken to be open.  Thus, as the strip is shifted in the perp-space direction by small amounts, exactly one of these two points is included in the projected set.  For any 1D lattice of points generated by $\vv$,  exactly one of these points will be included in the projected set.  

\begin{figure}
  \centering
  \includegraphics[width=\columnwidth]{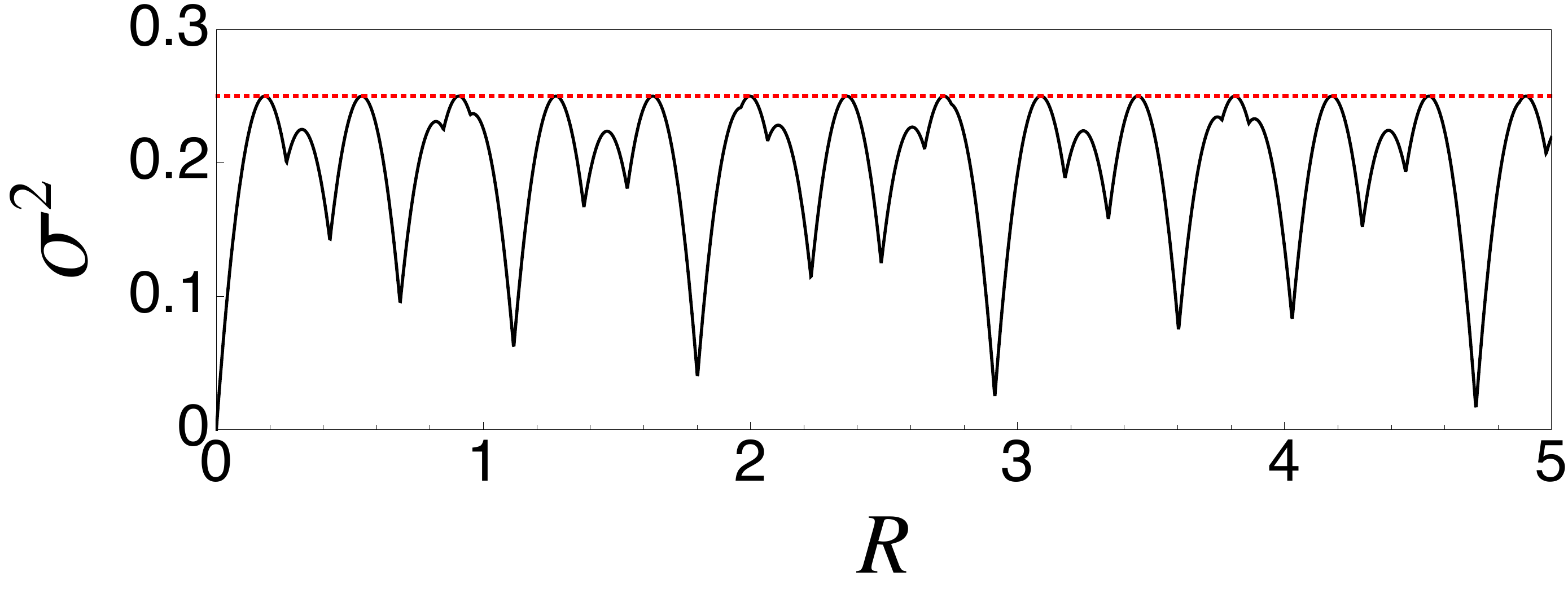}
  \caption{The analytically computed number variance for the canonical Fibonacci point set.  The dotted (red) line shows the upper bound of exactly $1/4$.}
\label{fig:sigma}
\end{figure}

Consider now a rectangular portion of the strip of length $R$, with $R \gg v_{\parallel}$, the physical space component of $\vv$.  As the rectangle is moved in the plane, any change in the number of points it covers must be due to points entering or leaving near the ends of the rectangle in the physical space.   As explained in detail in Appendix~\ref{app:ideal}, this permits the development of an exact analytic expression for $\sigma^2(R)$.  The result is that $\sigma^2(R)$ is a piecewise quadratic function that is bounded by zero from below and a constant of order unity from above.  
Figure~\ref{fig:sigma} shows an example for $\beta = \tau$.
The scaling law for $\sigma^2(R)$ is therefore trivial:
\begin{equation}
\sigma^2(R) \sim R^0\,,
\end{equation}
a result that is nicely consistent with Eq.~(\ref{eqn:alphanu}) and the above result $\alpha = 3$.

When $\omega$ is not of the form $i + j/\beta$, the above reasoning breaks down, and shifts in the position of the rectangle allow points to enter and leave asynchronously all along the length of the edges aligned with the physical space direction.  In this case, it is convenient to use an expression for $\sigma^2(R)$ involving a double sum over vectors of the 2D reciprocal space lattice, which must then be evaluated numerically.  The procedure is described in detail in Appendix~\ref{app:non-ideal}.  We find that the sum converges slowly; we must include more than $10^4$ terms in each of the sums in Eq.~(\ref{eq:sigmaexact}) to obtain accurate results.  The calculation clearly shows, however, that $\sigma^2(R)$ increases logarithmically with $R$.  This again is consistent with Eq.~(\ref{eqn:alphanu}) and the above result $\alpha = 1$.

\section{Discussion}

Our study of projected quasicrystalline point sets has both formal and practical implications.  One key result is the identification of the integrated spectral density $Z(k)$, rather than $S(k)$ or its envelope, as the quantity whose scaling behavior near $k=0$ determines the degree of hyperuniformity as measured by the scaling exponent $\alpha$.  The relation $Z \sim k^{1+\alpha}$ applies to quasicrystals as well as all previously studied structures.
Further, we find that the value of $\alpha$ for an important class of projected 1D quasicrystals 
depends on the width of the projection strip.  For ``ideal'' strips, we have $\alpha = 3$, while for non-ideal ones, $\alpha = 1$.  This observation establishes a new distinction between two classes of quasicrystalline point sets.  

Previous work established the connection between $\alpha$ and the number variance scaling exponent $\nu$.  In one dimension, $\nu=1$ for all $\alpha > 1$, but for $\alpha=1$ there is a logarithmic correction to $\sigma^2(R)$.  Our results confirm this connection for quasicrystals, with $\alpha$ determined from $Z(k)$.  Thus the difference in $\alpha$ between ideal ($\alpha=3$) and non-ideal strips ($\alpha=1$) has clearly observable consequences in the scaling of the number variance, suggesting that other physical properties may be differ between as well.  It would be interesting to study the nature of eigenstates or normal modes in these different classes of quasiperiodic structures.

\begin{figure}
  \centering
  \includegraphics[width=\columnwidth]{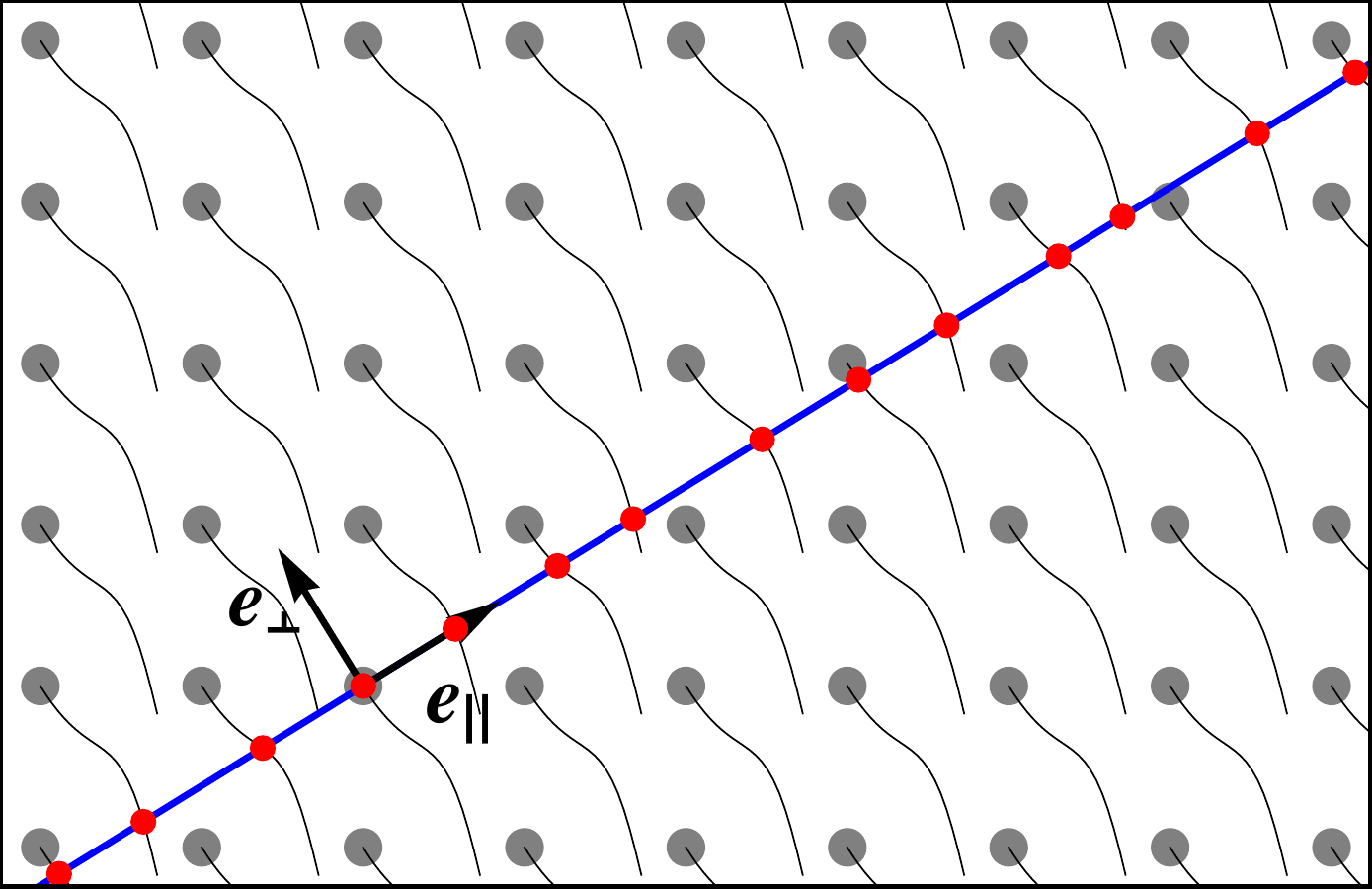}
  \caption{A quasicrystal generated as a cut through a hyperlattice of curved atomic surfaces.  The red points are the intersections of the curves and the physical line.}
\label{fig:cut}
\end{figure}

The present paper deals only with 1D quasicrystals projected from a 2D Bravais lattice.  Two types of generalization are straightforward.  First, one can decorate the hyperlattice unit cell with an arbitrary set of basis points without affecting $\alpha$ or $\nu$.  The decoration simply introduces a form factor in the Fourier transform of the hyperlattice, which modulates $S(k)$ but cannot change the scaling of $Z(k)$ as $k\rightarrow 0$, and it remains true that for the ideal case nearby points synchronously enter and leave the strip as it is shifted in the perp-space direction, implying that $\nu$ is not affected. 
Second, one can generalize the projection method to allow for ``curved atomic surfaces.''  Here each point in the hyperlattice is replaced by a surface (a curve when the perp-space is one-dimensional) and, rather than projecting the points within a strip, one takes the points where each curve intersects the physical space.  (See Fig.~\ref{fig:cut}.) In this case, the spacings between successive points generically take an infinite number of values rather than just two.  However, if the perp-space distance between the curve's endpoints is kept fixed, $\nu$ will not be affected by curvature in the segment; the number of points in a given interval of length $2R$ is the same as for the ordinary projected quasicrystal, with the possible exception of a bounded number of points at each end of that interval.

Other generalizations, including 1D quasicrystals projected from hyperlattices with dimension greater than 2 and higher-dimensional quasicrystals require further analysis.  Though some attention has been given to distinctions between structure factors of quasicrystals formed by decorations of the hyperlattice and decorations of tiles in physical space \cite{Jar86,Baa90}, we are not aware of any detailed studies of structures generated by non-ideal windows.  One may expect the distinction between ideal and non-ideal strip widths to arise in higher dimensions as well, but the calculations of $\alpha$ involve subtle effects that we have not yet addressed.

Finally, we note that ideal projected quasicrystals can be generated by substitution rules rather than projection \cite{Bom86}, which allows for a direct calculation of scaling exponents based only on the self-similarity of the structure. This approach can be generalized to substitution rules that yield qualitatively different types of spectra, including singular continuous and limit-periodic cases.~\cite{Aubry1988,Godreche1992,Godreche1989}  Our analysis of the scaling of $Z(k)$ and the hyperuniformity (or lack thereof) in 1D substitution sequences will be the subject of a future paper.


\newpage
\appendix
\section{Calculation of $S(k)$ for Fibonacci projection tilings} 
\label{app:Sk}
We wish to compute the structure factor $S(k)$ for a density consisting of a set of $\delta$-functions located at positions of the points on the physical line formed by projection of the subset of 2D lattice points that lie in a strip of width $w$ that is oriented with slope $1/\tau$.
For any irrational $\tau$, $S(k)$ can be obtained simply as the square of a convolution of the Fourier transform of the 2D lattice with the Fourier transform of $\Theta(\xv)$, where $\Theta(\xv)=1$ for $\xv$ in the strip and $0$ otherwise.  The transform of the lattice is, trivially, a set of $\delta$-functions at positions 
$(2\pi/a) (q \uvkx + p \uvky)$, with $p,q \in \mathbb{Z}$, where $\uvkx$ and $\uvky$ are the standard, orthogonal unit vectors in the lattice directions.  Rewriting $\uvkx$ and $\uvky$ in terms of unit vectors in the physical space and perp-space directions, $\kpara$ and $\kperp$, we have 
\begin{equation}
\kpara(p,q) = \frac{2\pi(p + q\tau)}{a\sqrt{1+\tau^2}}\,;\quad
\kperp(p,q) = \frac{2\pi(p\tau - q)}{a\sqrt{1+\tau^2}}\,.
\end{equation}
For notational convenience, we define $k_{pq}\equiv\kpara(p,q)$. 

The transform of $\Theta(\xv)$ is proportional to $\delta(\kpara) \sin(\kperp w/2) / (\kperp w/2)$.  Convolving this function  with the transform of the lattice and squaring to get peak intensities yields
\begin{equation}
S(k_{pq}) = C \left(\frac{\sin(w\,\kperp(p,q)/2)}{\kperp(p,q)}\right)^2\,
\end{equation}
where $C$ is a constant.
Using the identities 
\begin{equation}
\frac{a \tau}{\sqrt{1+\tau^2}}\kperp(p,q) = 2\pi p - \frac{a}{\sqrt{1+\tau^2}}k_{pq}
\end{equation}
and
\begin{equation}
k_{pq}\kperp(p,q) = \frac{(2\pi)^2\tau}{a^2(1+\tau^2)}(p^2-q^2+p\,q)\,,
\end{equation}
and defining $\omega$ such that 
\begin{equation}
w = \frac{a \tau}{\sqrt{1+\tau^2}}\,\omega\,,
\end{equation}
we find
\begin{equation}\label{eqn:generalS}
S(k_{pq}) = C' \left(\frac{(p+q\tau)\,\sin\left[\pi\omega \left(p - \frac{p+q\tau}{1+\tau^2}\right)\right]}{p^2-q^2+p\,q}\right)^2\,.
\end{equation}

\section{Calculations of $\sigma^2(R)$ for ideal windows} 
\label{app:ideal}
Let ${\cal Q}$ be the set of lattice points of a 2D square lattice with unit lattice constant; let $X$ be a line through the origin with slope $1/\beta$; and let ${\cal W}$ be a linear strip of width $w$ having $X$ as its lower (closed) boundary, where $w$ is chosen such that the upper (open) boundary of ${\cal W}$ passes through the lattice point $(-1,1)$; i.e., $w = (1+\beta)/\sqrt{1+\beta^2}$.  Define $\ev_{\parallel}$ and $\ev_{\perp}$ as the unit vectors along $X$ and orthogonal to $X$, respectively.  Note that $w = (-1,1)\cdot \ev_{\perp}$.  The set of points in $X$ is obtained by projecting all of the points in ${\cal Q}$ that lie within ${\cal W}$ orthogonally onto $X$.  (See Fig.~\ref{fig:projection}.)  In other words, the set of points of interest is  $\{(\xv\cdot\ev_{\parallel})\ev_{\parallel} \,\, | \,\,  0 \leq \xv\cdot\ev_{\perp} < w,\,\,\xv\in{\cal Q}\}$.

\begin{figure}
\centering
\includegraphics[width=\columnwidth]{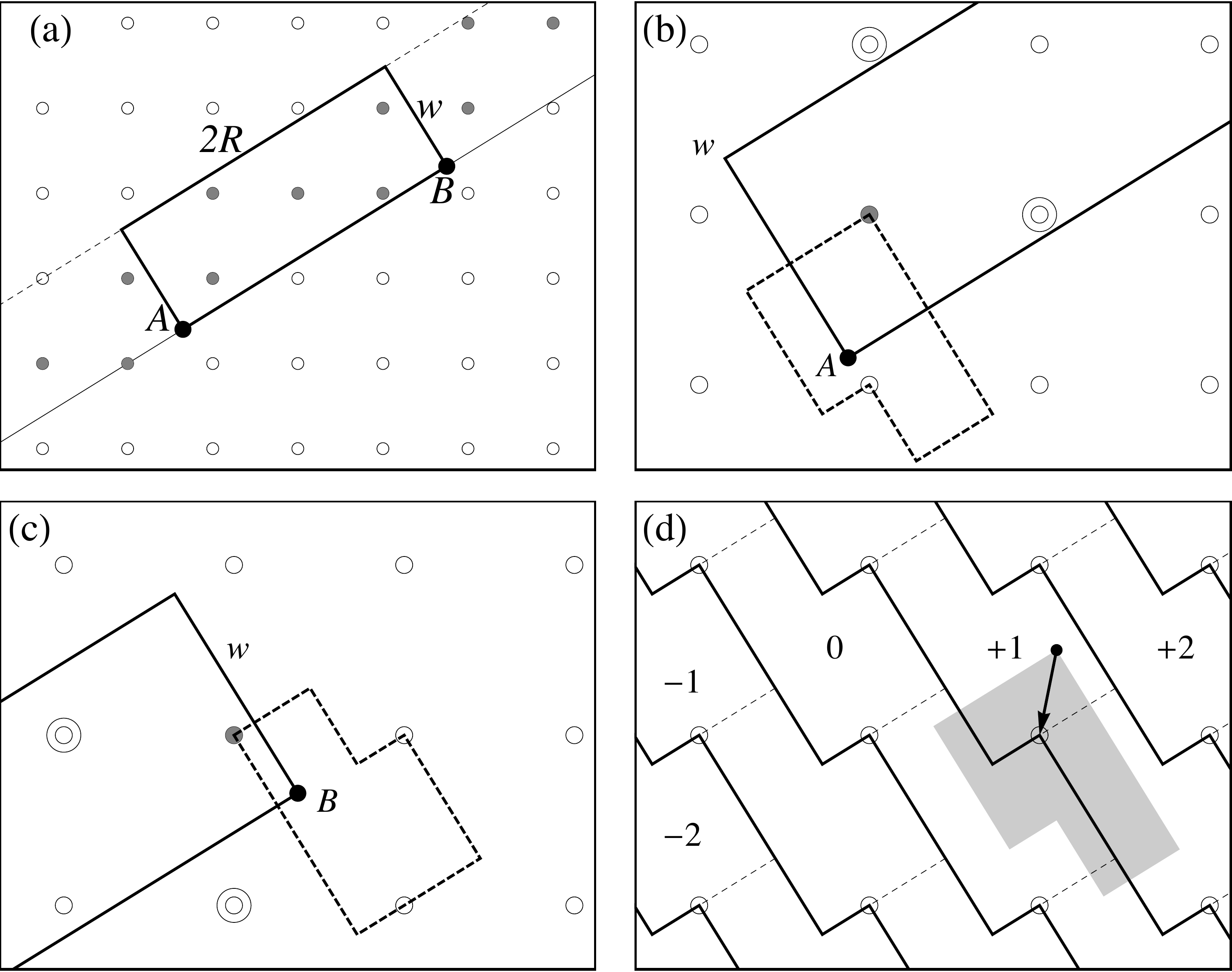}
\caption{Overlap areas for calculation of variances.  (a)  A portion of the projection strip showing one position of the window of length $R$.  (b) A view of one corner of the window.  The dashed region indicates where window corner $A$ must lie in order for the marked gray point to be the leftmost point in the window.  Exactly one of the doubly circled sites must be in the window for all positions of $A$ within the dashed region. (c)  The region in which the window corner $B$ must lie in order for the marked gray point to be the rightmost point in the window.  Exactly one of the doubly circled sites must be in the window for all positions of $A$ within the dashed region.  (d) The overlapping regions that determine the variance in the number of points within a finite strip.  The vector shown represents the relative displacement of $B$ with respect to $A$ modulo the lattice constant in both  the horizontal and vertical directions.  The numbers indicate the increasing number of points included in the strip for different locations of $B$.}
\label{fig:counting}
\end{figure}

We wish to compute the variance $\sigma^2(R)$ in the number of points on $X$ covered by a line segment of length $2R$ for random locations  of the left endpoint of the segment along $X$.  We assume for now that $\beta$ is an irrational number.  Our strategy is based on the geometry illustrated in Fig.~\ref{fig:counting}.
Panel (a) shows the projection strip ${\cal W}$ and a finite portion of length $2R$ having corners $A$ and $B$.  We refer to this rectangle as $W$.  Panel (b) show the region surrounding $A$.  As $W$ moves along ${\cal W}$, the position of $A$ within the unit cell uniformly covers the unit cell.  If $A$ lies anywhere within the dashed region, the point marked with a gray disk will be the leftmost point covered by $W$.  Similarly, panel (c) shows the region in which $B$ must lie in order for the gray point to be the rightmost one in $W$.  In both cases, the number of points within $W$ remains fixed for all locations of $A$ (or $B$) within the dashed region, with the possible exception of points at the other end of $W$; exactly one of the doubly circled pair of points must be included and similarly for all other pairs separated by the diagonal of the unit cell  along the length of $W$. 

Panel (d) shows the basis for the calculation of the variance for a given $R$.  The jagged lines demarcate regions with different numbers of points included in $W$ as $B$ is moved while $A$ is held fixed.  The ``0'' region is a reference for computing the variance, as we are not interested in the absolute number of points in $W$.   We refer to the region labeled by $n$ as $B_n$.  

Let $\rv(R)$ be the displacement of $B$ from $A$ modulo the basis vectors of the 2D lattice, indicated by an arrow in the figure:
\begin{equation} \label{eqn:rv}
\rv = \left(\{2R\, \ev_{\parallel,x}\},\{2R\, \ev_{\parallel,y}\}\right)\,,
\end{equation}
where $\{\cdot\}$ indicates the fractional part.  
A copy of the dashed region in Fig.~\ref{fig:counting}(b) is placed with its vertex at the base of the arrow, as shown in gray.  We refer to this region as $A_R$.  Note that $A_R$ exactly spans one unit cell of the lattice, and that all points within it correspond to one particular point being the leftmost in $W$.

A point within the gray region in Fig.~\ref{fig:counting}(d) represents a possible location of $A$, and the region it falls in gives the number of points in $W$ relative to the reference value.  Let $h(\rv,n)$ be the overlap area of $A_R$ and $B_n$.  As any location within the $A_R$ is equally likely, and $A_R$ has unit area, the variance is
\begin{equation}
\sigma^2(\rv) = \sum_n n^2 h(\rv,n) - \left( \sum_n n\, h(\rv,n) \right)^2\,.
\end{equation}
All that remains is to calculate the functions $h(\rv,n)$ for all $\rv$ within the unit cell.  It is clear from the geometry that all of the overlaps will be sums of rectangular areas, which will be quadratic functions of $x$ and $y$, the horizontal and vertical components of $\rv$.  Note that the calculation is trivial when $R=0$, as $A_R$ then falls entirely within $B_0$ and the variance is therefore zero.

\begin{figure}
\centering
\includegraphics[width=\columnwidth]{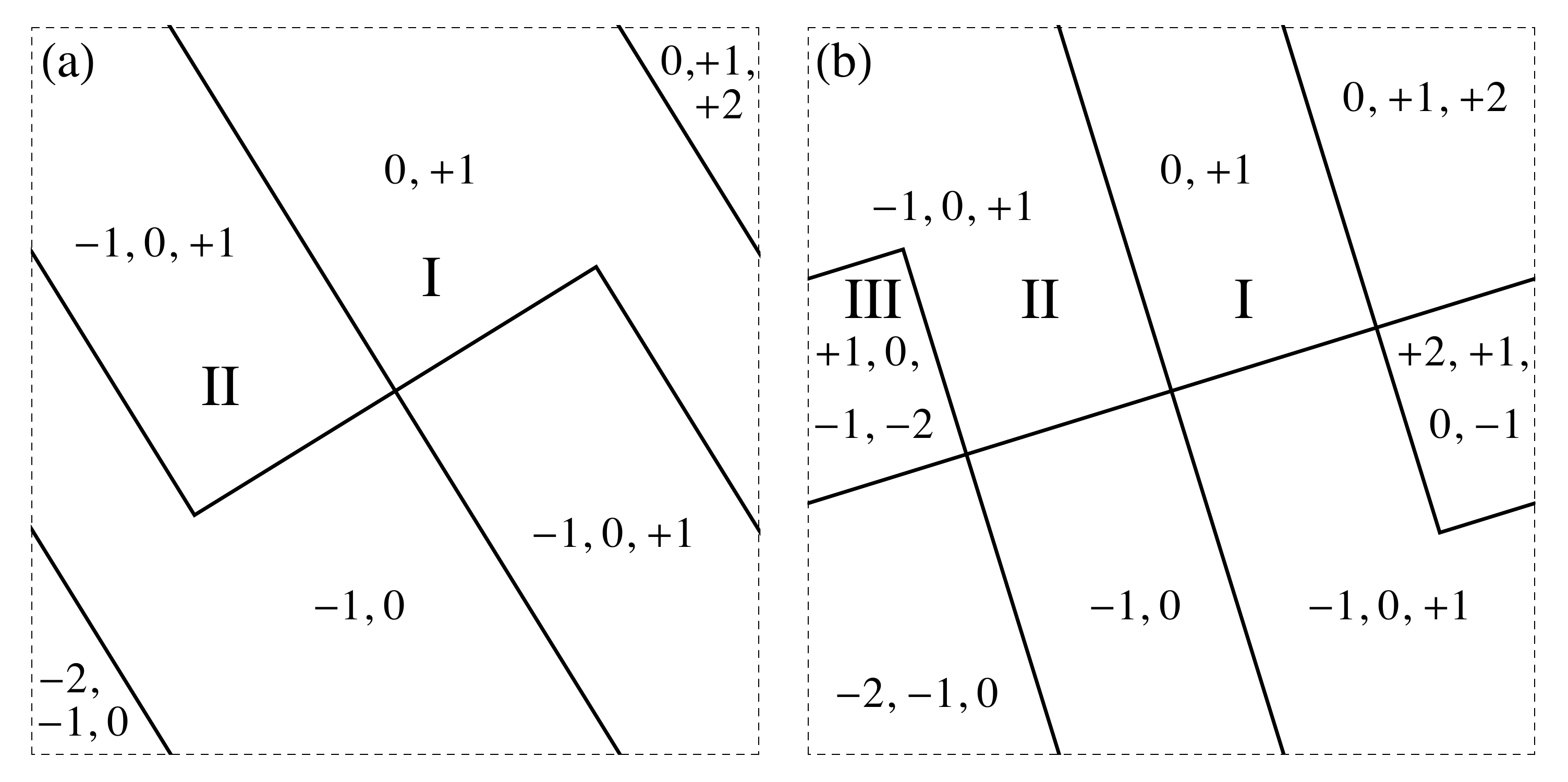}
\caption{Partition of the unit cell into regions with different overlap functions. (a) $\beta < 2$.  (b) $\beta > 2$.}
\label{fig:regions}
\end{figure}

$\sigma^2 (x,y)$ is a continuous, piecewise quadratic function with coefficients that change when a shift in $A_R$ causes it to overlap with a new region $B_n$.  There are two cases that must be handled separately, as shown in Fig.~\ref{fig:regions}.  Fig.~\ref{fig:regions}(a) shows the situation for $1 < \beta < 2$.
The unit cell is divided into six regions.  There is, however, an inversion symmetry corresponding to exchanging the roles of $A$ and $B$, as well as symmetry under translation by a lattice constant.  Thus it is sufficient to compute the overlap functions for regions I and II.  We take the unit cell to be bounded by $\pm 1/2$ in both directions.  After some algebra, we find for region I:
\begin{equation} \label{eqn:region1}
\sigma^2(\rv) = (x+y)(1-x-y)\,;
\end{equation}
and for region II:
\begin{align}\label{eqn:region2}
\sigma^2(\rv) = & -(x+y)^2-\left(\frac{\beta^2-1}{1+\beta^2}\right)(x-y-2xy) \nonumber \\ 
& - \left(\frac{2\beta}{1+\beta^2}\right) (x+y)(x-y+1)\,.
\end{align}

\begin{figure}[b]
\centering
\includegraphics[width=\columnwidth]{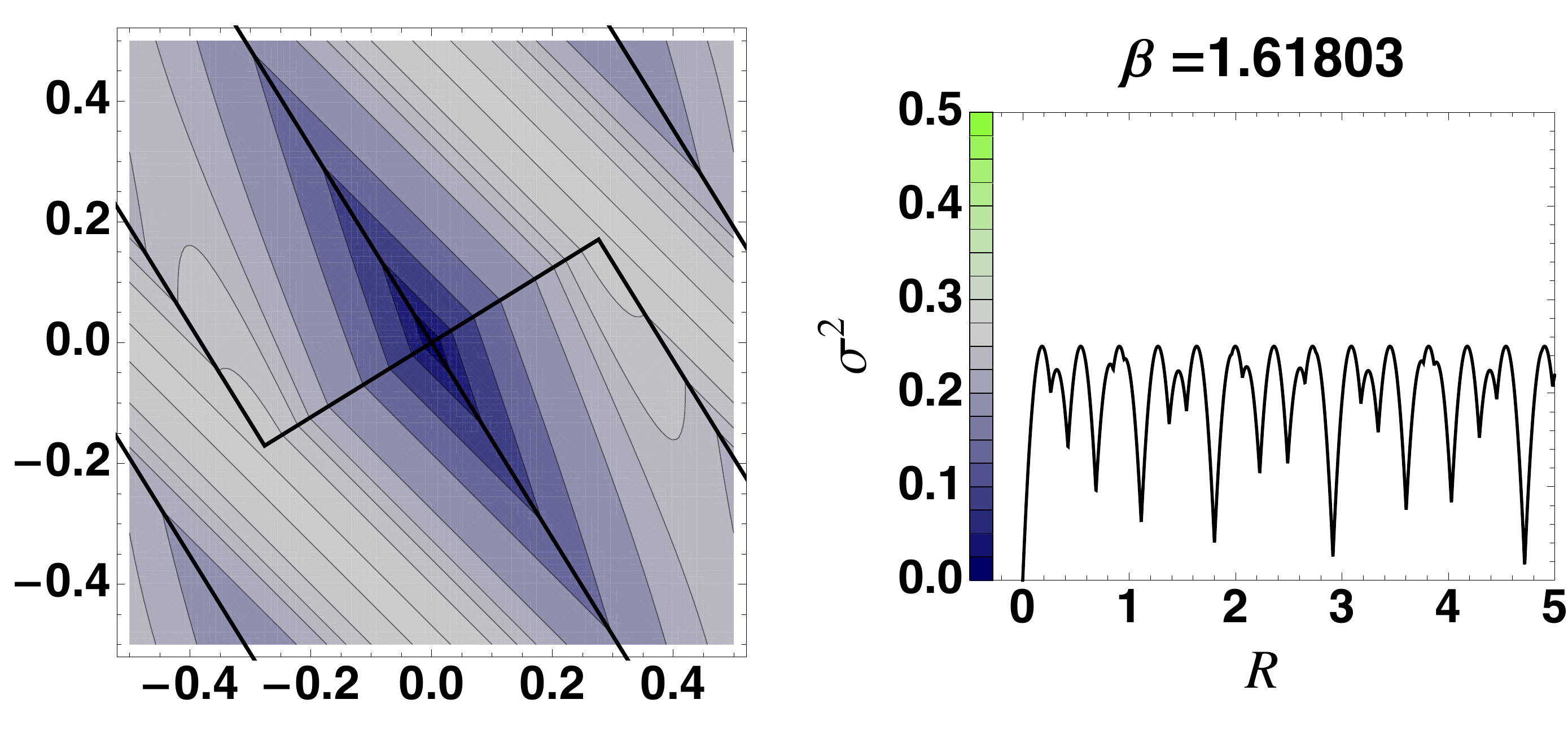}
\caption{Left: Contour plot of the variance function on the unit cell for $\beta = (1+\sqrt{5})/2$ (the golden mean).   Contour line values are not uniformly spaced.  The color bar shows a linear scale.  Right: The variance as a function of $R$, with $R$ measured in units of the 2D lattice constant.}
\label{fig:contours1}
\end{figure}

\begin{figure}
\centering
\includegraphics[width=\columnwidth]{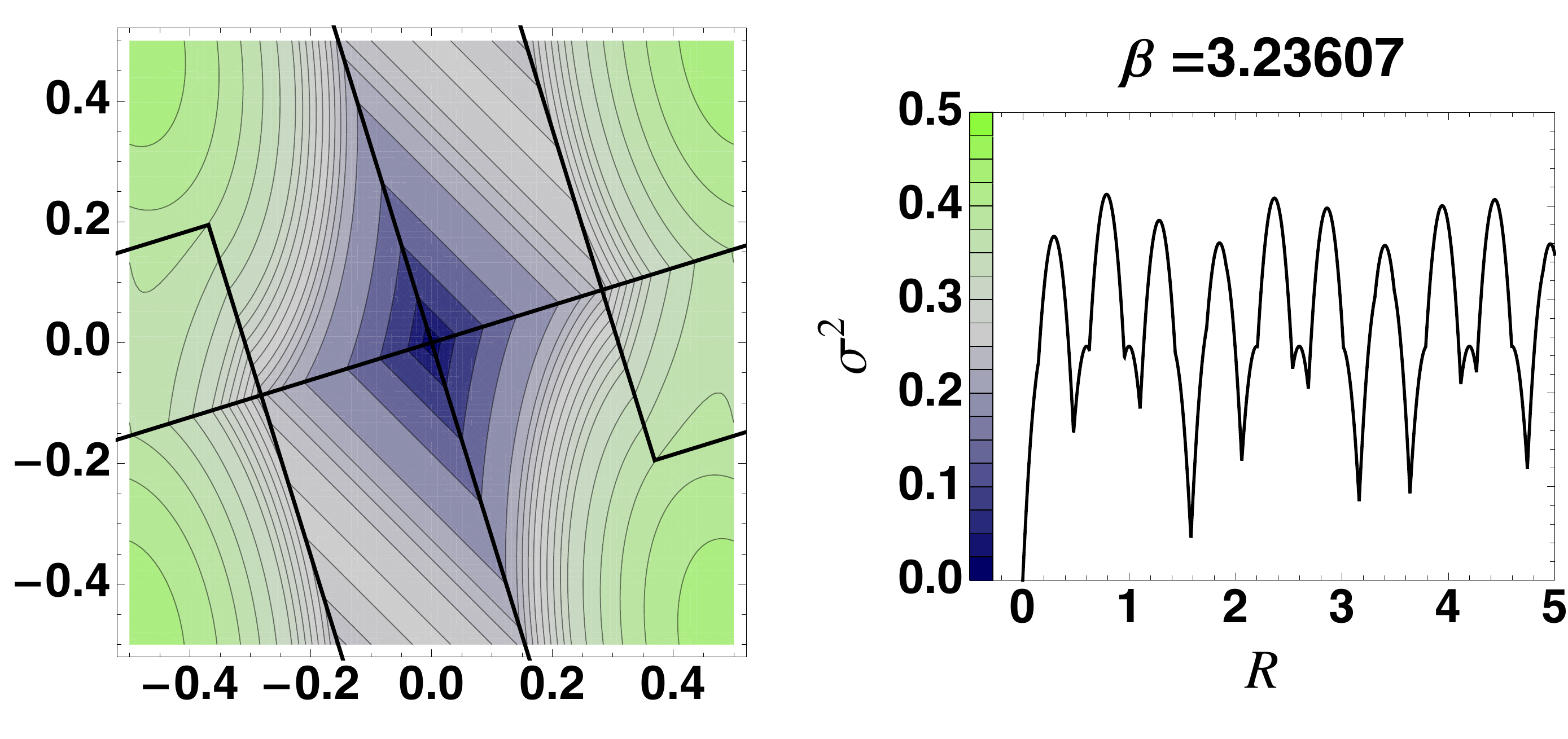}
\caption{Left: Contour plot of the variance function on the unit cell for $\beta = 1+\sqrt{5}$ (twice the golden mean).   Contour line values are not uniformly spaced.  The color bar shows the same linear scale as in Fig.~ref{fig:countours1}.  Right: The variance as a function of $R$, with $R$ measured in units of the 2D lattice constant.}
\label{fig:contours2}
\end{figure}

Fig.~\ref{fig:regions}(b) shows the situation for $2 < \beta$.  Here we need to compute overlaps for the three distinct regions marked in the figure.  The results for regions I and II are again given by Eqs.~(\ref{eqn:region1}) and~(\ref{eqn:region2}).   For region III, we find:
\begin{align}
\sigma^2(\rv) = & -(x-y)(x-y-1) \nonumber \\
& -\left(\frac{4\beta}{1+\beta^2}\right)(x^2-y^2+x) \nonumber \\ 
& -\left(\frac{2}{1+\beta^2}\right) (1+2y + 4xy)\,.
\end{align}
Contour plots of $\sigma^2(\rv)$ for the two cases are shown in Figs.~\ref{fig:contours1} and~\ref{fig:contours2}.  Note the simple ridge structure in region I, visible as straight lines in both cases.  The maximum value along the ridge is exactly $1/4$.  Note also the peak at $(x,y) = (-1/2,1/2)$ in region II.  From Eq.~(\ref{eqn:region2}) we find the value at the peak to be $(1/2)(\beta^2-1)/(\beta^2+1)$, which approaches $1/2$ for large $\beta$.  To obtain the plots of the variance as a function of $R$, we evaluate $\sigma^2(\rv)$ at the position dictated by Eq.~(\ref{eqn:rv}).

For rational values of $\beta$ it is no longer true that points $A$ and $B$ cover the unit cell uniformly as $W$ is translated along ${\cal W}$.  Nevertheless, shifting ${\cal W}$ in the $\ev_{\perp}$ direction does not change the sequence of points at all until the upper and lower boundaries of ${\cal W}$ both cross new lattice points, at which point the sequence shifts to a different locally isomorphic one.  Thus the averaging over the full unit cell still properly gives equal weight to all window positions.  The function $\rv(R)$ does not pass through all of the points in the unit cell, however, so that only a 1D subset of the values of $\sigma^2(\rv)$ are realized as $R$ increases.

Thus far we have shown that $\sigma^2(R)$ is bounded above by the highest peak in $\sigma^2(\rv)$ for a specific choice of $w$.  For the Fibonacci case, this is consistent with the result $\alpha = 3$ for $\omega=i+j/\tau$.  It is also consistent with expectations for a crystal when $\beta$ is rational.  The calculation of $\sigma^2$ for arbitrary $\beta$ shows that the behavior of $\sigma^2(R)$ is qualitatively similar for all $\beta$ and not dependent on the special properties of $\tau$ used in the calculation of $\alpha$, but because any $\alpha > 1$ results in the same scaling of $\sigma^2(R)$, we cannot conclude that all values of $\beta$ give $\alpha = 3$.

Extension of this analysis to the general case of $\omega=i+j/\beta$ is straightforward in principle.  Consider an arbitrary decoration of the unit cell of the 2D lattice; i.e., a lattice with a basis.  The analysis described above can be carried out in exactly the same way, the only difference being that there will be more boundary lines in Fig.~\ref{fig:counting}(d) and hence more distinct regions within the unit cell in Fig.~\ref{fig:regions}.  Thus $\sigma^2 (\rv)$ will still be a piecewise quadratic function that has the periodicities of the hyperlattice, though the number of pieces will increase with the number of points in the basis.  For any $\beta$, as long as $w$ is chosen such that the upper boundary of ${\cal W}$ passes through some lattice point, we can shear the lattice to map that point into $(-1,1)$ and thereby reduce the problem to that of a unit cell decorated with a finite number of points (and a different value of $\beta$).  The shear induces an affine transformation of the parallel space, which simply rescales $R$, while $\sigma^2 (R)$ remains a periodic, piecewise quadratic function.  These values of $w$ correspond precisely to values of $\omega$ of the form $i+j/\beta$, which is again consistent with the result above showing $\alpha = 3$ for the generalized Fibonacci case.

\section{Calculations of $\sigma^2(R)$ for non-ideal windows} 
\label{app:non-ideal}
We present here a method for numerically computing $\sigma^2(R)$ for non-ideal windows, in which case the upper boundary of ${\cal W}$ never passes through a lattice point.  In such cases the calculation in Appendix~\ref{app:ideal} breaks down because we cannot find pairs of doubly circled points like those in Fig.~\ref{fig:counting}(b) or~(c) that synchronously enter and leave ${\cal W}$.  As the window is shifted in the $\ev_{\perp}$ direction, points enter and leave asynchronously in the interior of the segment of length $2R$, making contributions to the variance that are not captured by the analysis of the changes occurring at the ends of the segment.  To treat this case, we develop an expression for $\sigma^2(R)$ as a double Fourier sum.  Note that the calculation of $Z(k)$ gives $\alpha = 1$, which predicts $\sigma^2(R) \sim \ln R$, a qualitatively different behavior than the previous case.

\begin{figure}
\centering
\includegraphics[width=0.8\columnwidth]{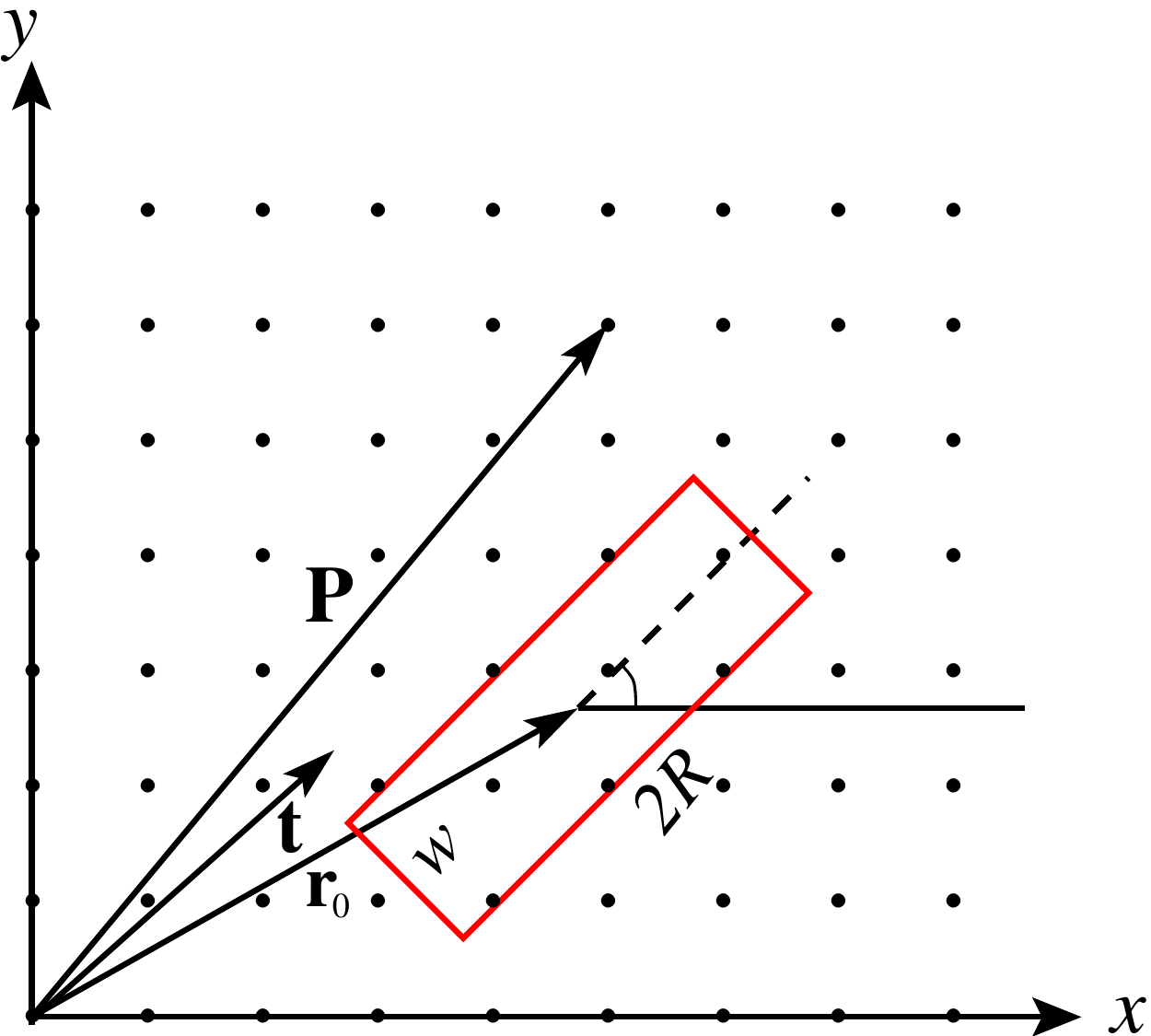}
\caption{Schematic model of number variance calculations of a subset of points of a square lattice. The number variance expression has been derived  for points in a rectangular window of width $w$ and length $2R$ as this window moves along the direction parallel to $R$ indicated by the dashed line.}
\label{fig:model}
\end{figure}

Consider a rectangular window of length $2R$ and width $w$ with the centroid at $\rv_0$, as shown in Fig.~\ref{fig:model}.
The number of points $N(\rv_0 ; R,w)$ within this window can be written as
\begin{equation}
N(\rv_0 ; R,w) = \sum_{\Pv} \Theta (R-|t_x|) \Theta (w/2-|t_y|),
\end{equation}
where $\Theta$ is the Heaviside step function, $\Pv$ the lattice vector, and $\tv = (t_x,t_y) = \Av (\Pv - \rv_0)$, with
$\Av$ denoting the rotation matrix (clockwise) in the plane. 

For irrational slopes of the window, averaging uniformly over all window positions is equivalent to averaging uniformly over the positions along the physical space line.  In this case, one can take advantage of the fact that
$N(\rv_0 ; R,w)$ is a periodic function in the window
position $\rv_0$ to write
\begin{equation}
\sigma^2(R) =  \frac{1}{v_c} \int_U \left[N(\rv_0; R,w)-\frac{2Rw}{v_c}\right]^2 d\rv_0,
\end{equation}
$v_c$ is the area of one unit cell of the lattice, $\int_U$ indicates an integral over one unit cell, and the subtracted constant is the average number of points in the window.
Expanding the integrand in a Fourier series gives \cite{To03a}
\begin{equation}
N(\rv_0 ; R,w) -\frac{2Rw}{v_c} =  \sum_{\kv \ne 0} b(\kv) e^{-i\kv\cdot\rv_0},
\end{equation}
where $\kv = (k_x, k_y)$ is a reciprocal lattice vector.
The Fourier coefficients are
\begin{eqnarray}
\nonumber  b (\kv)  &=& \dfrac{1}{v_c} \int_{U} N(\rv_0; R,w) e^{-i\kv\cdot\rv_0} d\rv_0 \\
\nonumber  &=& \dfrac{1}{v_c} \int_{\mathbb{R}^2} \Theta ( R - |t_x| ) \Theta ( w/2 - |t_y| ) 
		e^{-i \kv\cdot\Av\cdot\tv} d\tv \\
           &=& \dfrac{2Rw}{v_c} \mathrm{sinc} ( \kpara R) \mathrm{sinc} ( \kperp w / 2 ) ,
\end{eqnarray}
where we have used 
$d\rv_0 = |\det(-\Av^T)| d\tv = d\tv$ and 
\begin{align}
 \kpara & =    \cos(\phi)k_x + \sin(\phi) k_y  \,, \nonumber \\
 \kperp & =  - \sin(\phi)k_x + \cos(\phi) k_y  \,.
 \end{align}
Here, $\phi = \tan^{-1} (1 / \beta)$ indicates the tilt angle of the window with respect to $x$-axis.
Using Parseval's theorem, we can write the number variance as 
\begin{eqnarray}
\sigma^2  &=& \sum_{\kv \ne 0} b^2(\kv) \nonumber \\
         &=&  \left(\dfrac{2Rw}{v_c}\right)^2 \bigg[\!-\!1+  \label{eq:sigmaexact}\\
\nonumber         
         &\ & \quad\sum_{k_x=-\infty}^{\infty} \sum_{k_y=-\infty}^{\infty} \mathrm{sinc}^2 ( \kpara R ) \mathrm{sinc}^2 ( \kperp w/2 )\bigg]  \\
 &=& \left(\frac{4}{v_c}\right)\sum_{\kpara \neq 0} \frac{\sin^2 ( \kpara R ) \sin^2 (J(\kv) w/2\kpara )}{J^2(\kv)} \,,\label{eq:sigmaexact2}
\end{eqnarray}
where $J(\kv) = \kpara \kperp$.

For a non-ideal Fibonacci quasicrystal, we have shown $Z(k) \sim k^2$, i.e., $\alpha = 1$.  (See Fig.~\ref{fig:Znonideal}.)  From Eq.~(\ref{eqn:alphanu}) we thus expect the variance to scale as $\sigma^2 (R) \sim \ln R$.  The following rough argument shows how this comes about:  Consider a single scaling sequence of wavenumbers $\kpara$ given by $k_n = \kappa/\tau^n$ and the contribution it makes to the sum in  Eq.~(\ref{eq:sigmaexact2}).  For this sequence, the denominator $J^2$ is invariant, being proportional to the square of the invariant $I_{pq}$ of Eq.~(\ref{eqn:Ipq}).  For $n$ such that $k_n R \lesssim 1$, the first sine function in the numerator suppresses successive terms; the series of terms with $n > \ln(\kappa R)/\ln\tau$ converges.  Similarly, the second sine function suppresses terms with $n < -\ln(J w/2\kappa)/\ln\tau$.  For $n$'s between these two values, the terms are all generically of order unity in the non-ideal case (but not in the ideal case, by the same reasoning used for Eq.~(\ref{eqn:Skpq})  ), producing a sum of order $\ln R + \ln(J w/2)$ for large $R$.  This holds for each scaling sequence, with the factor of $1/J^2$ ensuring convergence in the sum over all scaling sequences.

\begin{figure}[h!]
  \centering
    \includegraphics[width=\columnwidth]{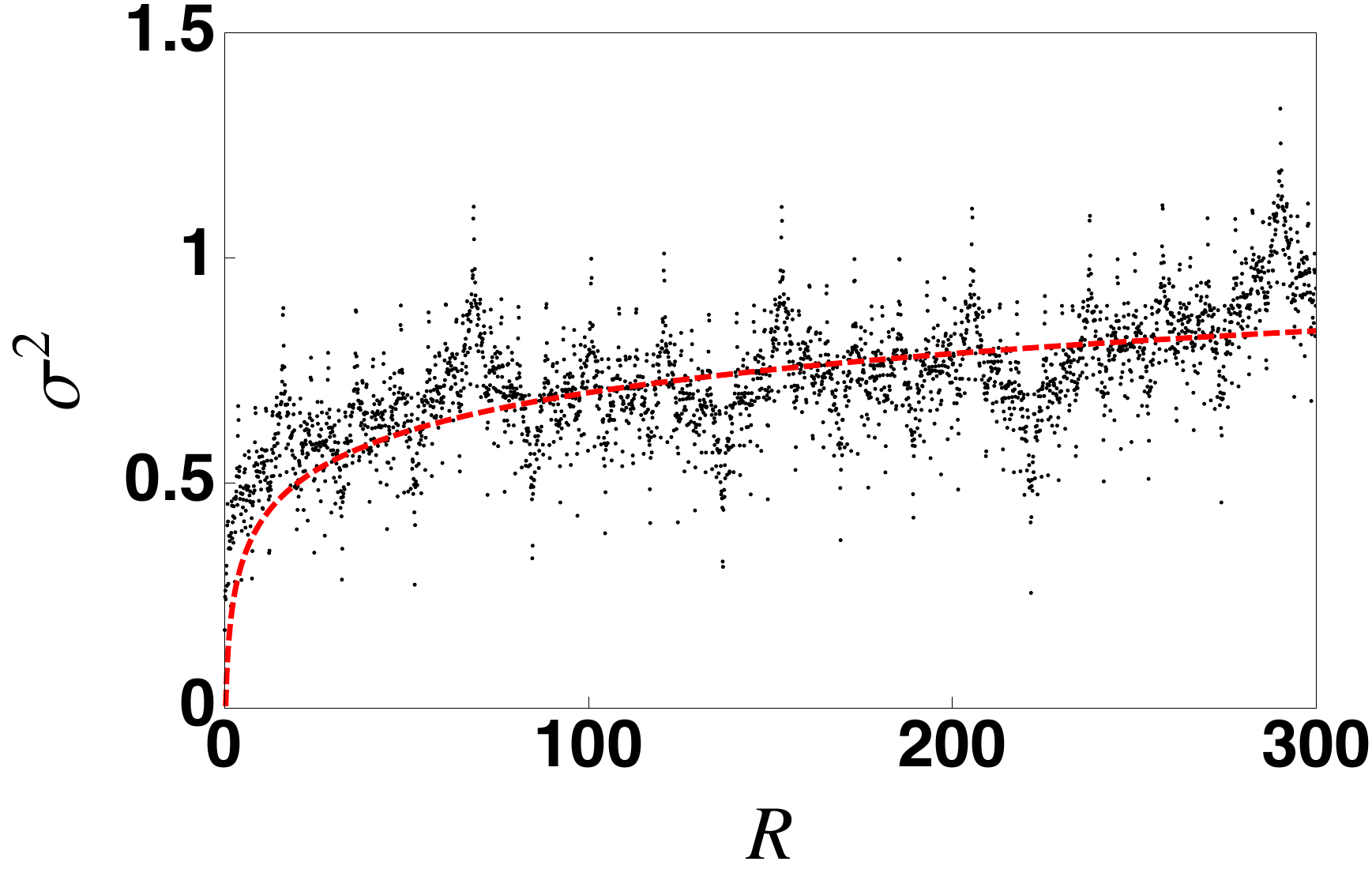}
      \caption{The number variance for a non-ideal Fibonacci quasicrystal as a function of $R$ and as obtained by Eq.~(\ref{eq:sigmaexact}). 
      		The red dashed line represents the function $(1+\ln R)/8$.  }
\label{fig:numvar_nonidFib}
\end{figure}

To verify this behavior, we evaluate the expression in Eq.~(\ref{eq:sigmaexact}) for $\omega = 1 / 4$. 
Figure~\ref{fig:numvar_nonidFib} shows the computed number variance as a function of $R$ for the non-ideal Fibonacci quasicrystal, and 
the logarithmic scaling for large $R$, indicated by the red dashed line, is confirmed.
The computed points include $10^4$ terms in each of the sums in Eq. ~(\ref{eq:sigmaexact2}).

\begin{figure}
  \centering
    \includegraphics[width=1.0\columnwidth]{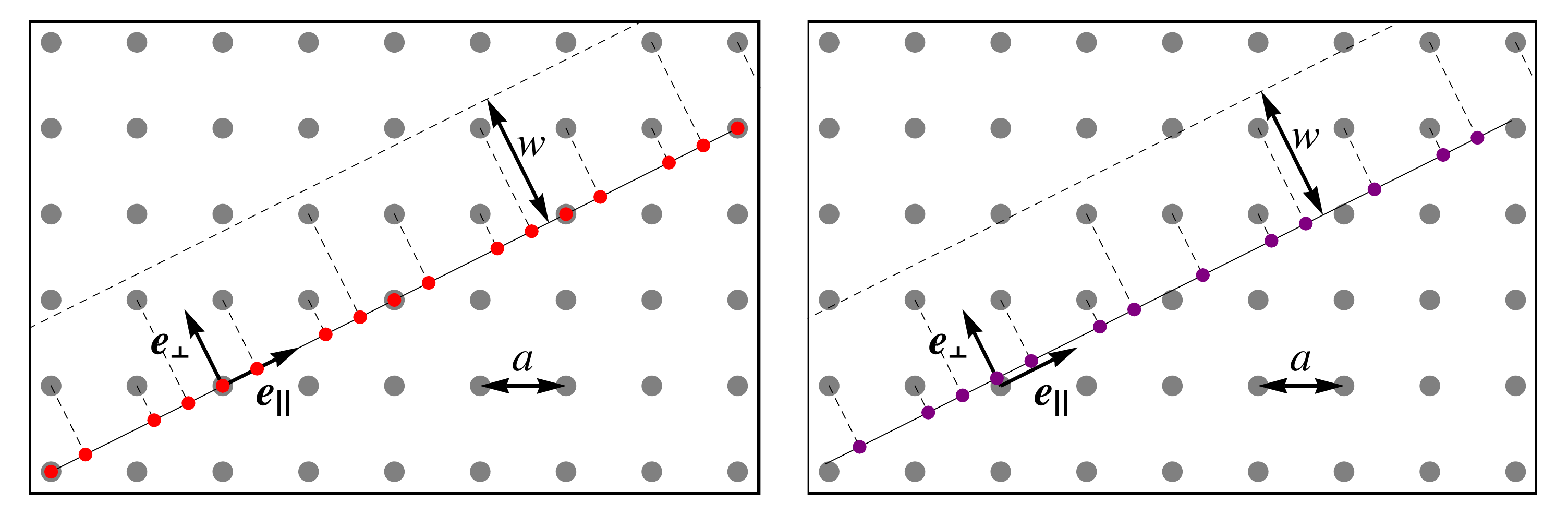}
      \caption{Two non-ideal rational windows of equal width that yield crystals with different densities.}
\label{fig:rationalwindows}
\end{figure}

Equation~(\ref{eq:sigmaexact}) applies whenever the projection is onto a line of irrational slope.  Care must be taken, however, in interpreting the results when applying it to rational projections.   For rational projections, we define ``ideal'' windows to be those for which the bottom (closed) boundary and top (open) boundary both pass through lattice points.  For ideal windows, all perp-space positions of the projection window yield the same crystal up to translation.  In this case, averaging over all window positions is equivalent to averaging over all parallel-space shifts of a given window, and Eq.~(\ref{eq:sigmaexact}) correctly gives $\sigma^2(R) \sim R^0$ for large $R$.  For non-ideal windows, on the other hand, different perp-space positions of the window can yield crystals with different densities, as shown in Fig.~\ref{fig:rationalwindows}.  In this case, averaging over all perp-space locations of the window yields $\sigma^2(R)\sim R^2$, even though any individual projected crystal must yield $\sigma^2(R)\sim R^0$.

\end{document}